\documentstyle[12pt]{article}

\input master.ipt
\input masterxt.ipt

\def\numt{\mbox{$\nu_{\mu(\tau)}$}} 

\def\pp{\mbox{pp}}
\def\pep{\mbox{pep}}
\def\CNO{\mbox{CNO}}
\def\Ps{\mbox{${\bar{P}}_\odot$}}
\def\PTot{\mbox{$P_\oplus$}}
\def\APTot{\mbox{${\bar{P}}_\oplus$}}

\def\CdTv{\mbox{$\cos 2 \theta_V$}}
\def\CdTM{\mbox{$\cos 2 \theta_m$}}

\def\lambdaDtc{\mbox{$\lambda_D$}}

\def\NA{Nadir angle}

\def\TYear{\mbox{$T_y$}}
\def\RSE{\mbox{$R$}}

\def\ltap{\lsim}  

\def\FrameP{(a)} 
\def\FrameN{(b)} 
\def\FrameC{(c)} 
\def\FrameM{(d)} 

\def\rescaled{\mbox{$\APeTw \times 10$}}   

\def\FigOne{1}                   
\def\FigEarthEffectGeometry{1}   
\def\FigPeTw{2}                  
\def\FigAPeTw{3}                 
\def\FigForest{4}                

\def\TabOne{I}
\def\TabActiveNeutrinoDots{I}     
\def\TabResidenceTime{II}         

\def\Item{}
\def\BeginE{}
\def\EndE{}

\begin{document}
\sloppy

%
%
{\normalsize
\begin{flushright}
\begin{tabular}{l}
Ref. SISSA 16/97/EP \\
January 1997\\
arch-iv/9702361
\end{tabular}
\end{flushright}

\vspace{0.5cm}


\begin{center}
{\Large
A Study of the Day - Night Effect for the \SK\ \\
\vspace{0.25cm}
Detector: I. Time Averaged Solar Neutrino Survival Probability 
}  
\end{center}

\begin{center}
Q.Y. Liu $^{\mbox{a)}}$, 
M. Maris $^{\mbox{a,b)}}$, 
S.T. Petcov 
\footnote{
Also at: Institute of Nuclear
Research and Nuclear Energy, Bulgarian Academy of Sciences,
1784 Sofia, Bulgaria.}
$^{\mbox{a,c)}}$\\
$^{\mbox{a)}}$ Scuola Internazionale Superiore di Studi Avanzati, 
Trieste, Italy.\\
$^{\mbox{b)}}$ INFN - Sezione di Pavia, Pavia, Italy.\\
$^{\mbox{c)}}$ INFN - Sezione di Trieste, Trieste, Italy.
\end{center}


\vspace{1cm}

\bec
\abstract{
\noindent
This is the first of two articles 
aimed at providing comprehensive predictions for the day-night (D-N) 
effect for the Super-Kamiokande detector in the case of the MSW
$\nue \rightarrow \numt$ transition solution of the solar 
neutrino problem. The one-year averaged probability of survival
of the solar $\nue$ crossing the Earth mantle, the core, the inner 2/3 of the 
core, and the (core + mantle) is calculated with  
high precision (better than 1\%)
using the elliptical orbit approximation (EOA) to describe the
Earth motion around the Sun. Results for the survival probability in the
indicated cases are obtained for a large set of values of the MSW
transition parameters $\Delta m^2$ and $\sin^22\theta_{V}$ from the
``conservative'' regions of the MSW solution, derived by taking into account
possible relatively large uncertainties in the values of the $^{8}$B and
$^{7}$Be neutrino fluxes. Our results show that the one-year averaged
D-N asymmetry in the $\nue$ survival probability for neutrinos crossing the
Earth core can be, in the case of $\sin^22\theta_{V} \leq 0.13$, 
larger than the asymmetry in the probability
for (only mantle crossing + core crossing) neutrinos
by a factor of up to six. 
The enhancement is
larger in the case of neutrinos crossing the inner 2/3 of the core.
This indicates that  
the Super-Kamiokande experiment might be able to 
test the $\sin^22\theta_{V} \leq 0.01$
region of the MSW solution of the solar neutrino problem
by performing selective D-N asymmetry measurements.    
}
\eec

\newpage
\section{Introduction}
\BeginE
\Item 
The Earth effect in neutrino propagation is a direct consequence of 
neutrino oscillations in matter \cite{MSW:Original}. 
If the neutrino mass spectrum is not degenerate and if neutrino mixing takes place in 
vacuum, the neutrino propagation is sensitive to the matter distribution along
the propagation path and the probability to detect at Earth a \nue\ 
produced in the Sun is a function of the detection time because at night the Sun is 
below the horizon and the solar 
neutrinos cross the Earth reaching the detector. 

\Item 
The observation of the Earth, or day-night (D-N), effect would be a proof 
of the validity of the MSW solution of the solar neutrino problem.
In its simplest version the MSW mechanism involves matter - enhanced 
two-neutrino transitions of the solar \nue\ into an active neutrino 
\num\ or \nut, $\nue \rightarrow \nu_{\mu(\tau)}$, while the \nue\ 
propagates from the central part to the surface of the Sun.
The $\nue \rightarrow \nu_{\mu(\tau)}$\ transition probability depends in 
this case on two parameters: \dms\ and \SdTvS, where $\dms > 0$\ is 
neutrino mass squared difference and $\ThetaV$\ is the angle 
characterizing the neutrino mixing in vacuum.
When the $\nu_{\mu(\tau)}$\ produced in the Sun due to the MSW transitions
and the solar \nue\ not converted into \numt\ in the Sun cross the Earth,
further $\nue \rightarrow \numt$\ transitions and/or \nue\ regeneration 
due to the inverse MSW process $\numt \rightarrow \nue$\ can take place. 
This can lead to a difference between the signals caused by the solar 
neutrinos in a solar neutrino detector during the day and during the night, 
i.e., to a D-N asymmetry in the signal.
No other mechanism of depletion of the solar \nue\ flux proposed so far 
can produce such an effect.

\Item
The MSW solution of the solar neutrino problem and the D-N effect related 
to it have been extensively studied.
The latest solar neutrino data can be described for values of the parameters
\dms\ and \SdTvS\ belonging to the intervals \cite{Krastev:Petcov:1996}
\bec\beq\label{eq:kpnu96:a}
  \begin{array}{c}
  3.6 \times 10^{-6} \, \mbox{ eV}^2 \, \ltap \, \Delta m^2 \, \ltap \, 
9.8\times 10^{-6}\mbox{ eV}^2,\\
  \\
  4.5 \times 10^{-3} \, \ltap \, \SdTvS \, \ltap \, 1.3\times 10^{-2}, \\
  \end{array}
\eeq\eec

\noindent
or

\bec\beq\label{eq:kpnu96:b}
  \begin{array}{cc}
    5.7 \times 10^{-6} \mbox{ eV}^2 \, \ltap \, \Delta m^2 \,\ltap\, 9.5\times 10^{-5} \mbox{ eV}^2,\\
    \\
    0.51 \,\ltap \, \SdTvS \, \ltap\, 0.92, \\
  \end{array} 
\eeq\eec

\noindent
assuming the solar \nue\ MSW transitions are into active neutrinos. 
The intervals (\ref{eq:kpnu96:a}) and (\ref{eq:kpnu96:b})
correspond to the nonadiabatic (small mixing angle) and to the adiabatic
(large mixing angle) solutions of the solar neutrino problem.
They have been obtained using the predictions of 
the solar model of Bahcall and Pinsonneault from 1995 
\cite{Bahcall:Pinsonneault:1995}\ for the fluxes of the different
solar neutrino flux components: \pp, \pep, \BeSv, \BHt\ and \CNO. 
This model takes into  account heavy elements diffusion and was recently shown to be in 
very good agreement with the latest and more precise helio-seismological observations 
\cite{Bahcall:Pinsonneault:etal:1996}.

\Item
The possible solution values of \dms\ and \SdTvS\ are important in the 
calculations of the D-N effect because they determine the magnitude of the
effect. 
A more conservative approach to the determination of the MSW solution 
regions of values of \dms\ and \SdTvS\ allows for, e.g., the spread in 
the predictions for the \BHt, \BeSv, etc. fluxes in the contemporary solar
models, the possible changes in the predictions associated with the 
uncertainties in the relevant nuclear reaction cross-sections which are used
as input in the flux calculations, 
etc.\footnote{Although the total fluxes of \BHt, \BeSv, etc. neutrinos 
depend on the physical conditions in the central part of the Sun, 
the spectra of the \BHt, \CNO\ and the \pp\ neutrinos are solar physics 
independent.}.
Such an approach \cite{MSW2}\ yields larger regions for 
both the nonadiabatic and the adiabatic MSW solutions \cite{Krastev:Petcov:1996}:
\bec\beq\label{eq:kpnu96:aII}
  \begin{array}{c}
  3.0 \times 10^{-6} \mbox{ eV}^2 \,\ltap\, \Delta m^2 \,\ltap\, 1.2\times 
10^{-5}\mbox{ eV}^2,\\
  \\
  6.6 \times 10^{-4} \,\ltap\, \SdTvS \,\ltap\, 1.5\times 10^{-2}, \\
  \end{array} 
\eeq\eec

\noindent
or

\bec\beq\label{eq:kpnu96:bII}
  \begin{array}{cc}
    7.0 \times 10^{-6} \mbox{ eV}^2 \,\ltap\, \Delta m^2 \,\ltap\, 1.6\times 
10^{-4} \mbox{ eV}^2,\\
    \\
    0.30 \,\ltap\, \SdTvS \,\ltap\, 0.94. \\
  \end{array} 
\eeq\eec

\noindent
It is hardly conceivable at present that the values of \dms\ and \SdTvS\ 
for which the MSW $\nue \rightarrow \numt$\ transition mechanism provides 
a solution of the solar neutrino problem can be very different from those
shown in Fig. \FigOne\ and given in equations (\ref{eq:kpnu96:aII}) and
(\ref{eq:kpnu96:bII}).

\Item
The D-N effect was widely studied
\cite{
Cribier:etal:1986, 
MS:DNEFF:1987,
Carlson:1986, 
Hiroi:etal:1987, 
Baltz:Weneser:1987,
Dar:Mann:1987,
Baltz:Weneser:1994,
Hata:Langacker:1993Earth,
Krastev:1996,
Gelb:Kwong:Rosen:1996}.
It was pointed out, in particular, that the effect can be significant 
for $\SdTvS > 0.01$\  for a large range of values \dms\ and \SdTvS\ from
the large mixing angle solution region, for which the Earth effect regenerates 
electron neutrinos increasing the event rate at night.
At the same time results from Kamiokande II and III experiments  
\cite{Kamiokande:Results}\ 
did not reveal any significant Earth effect exceeding approximately $30\%$,
thus excluding a rather large area of the region of values of \dms\ and 
\SdTvS\ of the large mixing angle solution \cite{Hata:Langacker:1993Earth}.
This area is already excluded by the existing mean 
event rate data from the Cl-Ar, Ga-Ge and Kamiokande experiments.

\Item 
The motivation for new quantitative theoretical studies of the D-N asymmetry 
comes from a number of considerations regarding the earlier studies and the 
experimental situation which is reasonable to expect in the near future.
First of all the discovery of a \daynight\ effect, even of a small one,
should be an important confirmation of the neutrino oscillation
hypothesis and will open a new chapter of physics, while the absence
of any detectable effect should represent a strong constraint on the neutrino 
mass and mixing parameters relevant to the MSW solution.
It is clear that in both cases quantitative predictions are required
to assure a proper interpretation of experimental results.
Second, most  of the previous studies were primarily concerned with the 
nature of the effect
\cite{Cribier:etal:1986,
Carlson:1986,
Hiroi:etal:1987,
Baltz:Weneser:1987}.
The relevant probability distributions and event rates were calculated  
for some representatives values of the parameters only, which is insufficient 
for a comprehensive understanding of the possible magnitude of the effect. 
Alternatively, in some of the articles on the subject the effect of the 
Earth on the oscillations of neutrinos was investigated for high energy
neutrinos produced by cosmic rays or accelerators and the results obtained
were extended to solar neutrinos 
\cite{Carlson:1986, Ermilova:Tsarev:Chechin:1986, Dar:Mann:1987,
LoSecco:1993} 
(see also, e.g., \cite{Krastev:Petcov:1988E}).
The possible application of the Earth effect to the problem of the 
reconstruction of the Earth internal structure (Earth Tomography) was 
also considered 
\cite{Ermilova:Tsarev:Chechin:1986, Nicolaidis:1988, Bertotti:Maris:1996}
(see also \cite{Carlson:1986, Hiroi:etal:1987, Dar:Mann:1987}). 
More detailed results on the Earth effect have been obtained in 
\cite{Baltz:Weneser:1994}\ where the D-N asymmetry for the \SK\ detector,
for instance, was calculated for 8 (4) sets of values of \SdTvS\ and \dms\ 
from the region (\ref{eq:kpnu96:a}) ((\ref{eq:kpnu96:b})) of the nonadiabatic 
(adiabatic) solution, as well as in \cite{Krastev:1996}\ where the one year 
average iso - (D-N) asymmetry contours in the \dms\ - \SdTvS\ plane for the 
\SK\ and SNO detectors have been derived.
Third, the Kamiokande experiment, like the other solar 
neutrino experiments which
have provided data so far, has a relatively low statistics.
This problem will be overcome by new experiments such as \SK, SNO and ICARUS.
They are expected to be much more sensitive to the D-N effect,
allowing to exploit a wider neutrino parameter region where 
the Earth effect is not large, in particular, the region of the small 
mixing angle nonadiabatic solution. One of them, \SK, is already operating and 
is expected to produce soon new results.
Finally, an improved statistics allows to develop more sophisticated 
tests enhancing the sensitivity to the effect.

%
%
%
%

\Item
The present article represents the first of two articles aimed at providing
comprehensive predictions for the D-N effect for the \SK\ detector
in the case of the MSW $\nue \rightarrow \numt$\ transition solutions of the
solar neutrino problem.
These include: i) calculation of the Earth effect with a sufficiently high
precision which can match the precision of the data on the effect to be 
provided by the \SK\ detector, ii) calculation of the magnitude of the effect 
for a sufficiently large and representative set of values of the 
parameters  from the ``conservative'' regions (\ref{eq:kpnu96:aII}) and 
(\ref{eq:kpnu96:bII}) of the MSW solution, iii) calculations of the one-year
averaged D-N asymmetry for solar neutrinos 
crossing the Earth mantle, the core, the inner 2/3 
of the core, and the mantle + core (full Earth) for the chosen large set of 
values of \dms\ and \SdTvS\ with the purpose of illustrating the 
magnitude of the enhancement of the effect which can be achieved by an 
appropriate selection of the data sample, 
iv) calculation of the one year average deformations of the $e^-$\ 
spectrum by the Earth effect in the mantle, core, and in mantle + core 
for the selected representative set of values of \dms\ and \SdTvS. 

\Item
The magnitude of the D-N effect is determined by the MSW probability of solar
\nue\ survival when the solar neutrinos cross the Earth to reach the detector,
$\PTot(\nue \rightarrow \nue)$, as well as by the probability of solar \nue\ 
survival in the Sun, $\Ps(\nue \rightarrow \nue)$, describing the MSW effect 
on the solar neutrino flux detected during the day. 
Thus, the first step in the realization of our program of studies of the D-N
effect consists of sufficiently precise calculation of the one year average 
probability $\APTot(\nue \rightarrow \nue)$\ in the cases of interest, i.e.,
for neutrinos crossing the mantle, the core, 2/3 of the core and the 
mantle + core when traversing the Earth.
In the present article this first step is accomplished.

\Item
\EndE

\section{Calculating the Earth Effect}\label{section:2}
\BeginE
%

\Item
Given that \SK\ detector should have an event rate of about $10^4$\ 
solar neutrino induced events per year (with a $5$ MeV energy threshold), 
the expected statistical accuracy in the flux measurement is about $1\%$. This 
means that the numerical accuracy in computing the Earth effect has to be 
better than approximately $1\%$\ each 
time the effect is not smaller than $1\%$.

\Item
Comprehensive and sufficiently accurate predictions require 
complex computations.
In order to obtain detailed predictions for the magnitude of the D-N effect 
and how it changes with the change of the values of the two 
parameters \dms\ and \SdTvS,
we have chosen to calculate the effect for a large 
set of points from the MSW solution
regions (\ref{eq:kpnu96:aII}) and (\ref{eq:kpnu96:bII}). The values of \dms\ and \SdTvS\ 
corresponding to these points are given in Table \TabOne. The points selected are 
distributed evenly in the ``conservative'' regions of the nonadiabatic and the 
adiabatic solutions (\ref{eq:kpnu96:aII}) and (\ref{eq:kpnu96:bII})
\cite{Krastev:Petcov:1996}. 

\Item 
We have used the Stacey model \cite{Stacey:1977}\ as a reference model for the 
Earth density distribution in our calculations. 
Obviously, this is not an unquestionable choice because new models were produced in the last years (see, e.g., \cite{Jeanloz:1990}). 
However, the merit of the Stacey model when compared with 
alternative ones is its ability to describe the known details of the Earth radial density 
distribution, $\rho(R)$, relevant for the Earth effect calculations. This is done through a set
of polynomials which allows the use of very efficient numerical computation schemes. 
In the Stacey model the Earth is assumed to be spherical and isotropical 
with a radius of 6371 km. As in all Earth models, there are two main density 
structures in the Stacey model: mantle and core. The core has a radius of 
3485.7 km and a mean density of approximately 11.5 gr/cm$^3$. The mantle 
surrounding the core has a mean density of about 4 gr/cm$^3$. Both of them 
have a number of density substructures. 

\Item
The Earth effect formalism was described by many authors and it was reviewed recently in
ref. \cite{Baltz:Weneser:1994}. The probability that an electron neutrino produced in 
the Sun will not be converted into \numt\ when it propagates in the Sun and 
traverses the Earth on the way to the detector is given by 
\cite{MS:DNEFF:1987}\
\bec\beq\label{eq:neutrino:survival:proba}
    \PTot(\nue \rightarrow \nue) = \Ps(\nue \rightarrow \nue) + 
             \frac{1 - 2\Ps(\nue \rightarrow \nue)}{\CdTv} 
      \, (\PeTw - \STvS),
\eeq\eec

\noindent
where $\Ps(\nue \rightarrow \nue)$\ is the average probability of solar \nue\ 
survival in the Sun and \PeTw\ is the probability
of the $\nutwo \rightarrow \nue$\ transition after the \nue\ have left the Sun.
During the day, when the neutrinos do not cross the Earth, $\PeTw = \STvS$\ and 
we have $\PTot(\nue \rightarrow \nue) = \Ps(\nue \rightarrow \nue)$. 
For Earth crossing neutrinos at night $\PeTw\neq\STvS$\ due to the MSW effect
and $\PTot(\nue \rightarrow \nue) \neq \Ps(\nue \rightarrow \nue)$.
Throughout this study we use for $\Ps(\nue \rightarrow \nue)$\ the expression derived 
in ref. \cite{Petcov:1988}\ in the exponential approximation for the variation of the
electron number density, $N_e$, along the neutrino path in the Sun:
\bec\beq\label{eq:Ps}
    \Ps(\nue \rightarrow \nue) = \frac{1}{2} + \left(\frac{1}{2} - P'\right)
         \CdTv \CdTM,
\eeq\eec

\noindent
where \ThetaM\ is the neutrino mixing angle in matter at the point of \nue\
production  in the Sun and 
\bec\beq
    P' = \frac{ e^{-2\pi r_0 {{\Delta m^2}\over{2E}}\sin^2 2 \theta_v} - 
                e^{-2\pi r_0 {{\Delta m^2}\over{2E}}}
              } 
              {
                 1 - e^{-2\pi r_0 {{\Delta m^2}\over{2E}}}
              }
\eeq\eec

\noindent
is the analog of the Landau-Zener jump probability for exponentially varying $N_e$, 
$r_0$\ being the scale height characterizing the change of $N_e$\ along the neutrino trajectory.
The probability $\Ps(\nue \rightarrow \nue)$, eq. (\ref{eq:Ps}), was averaged over the
region of \BHt\ neutrino production in the Sun. 
For values of $\SdTvS \, \lsim \, 4 \times 10^{-3}$\ the probability  
$\Ps(\nue \rightarrow \nue)$ 
was calculated following the prescriptions given in \cite{Krastev:Petcov:1988}.

\Item
Let us discuss next the calculation of the probability \PeTw. 
In contrast to $\Ps(\nue \rightarrow \nue)$, the probability \PeTw\ is not 
a constant in time because it is a function of the trajectory followed by the 
neutrinos crossing the Earth. The latter is determined by the instantaneous 
apparent position of the Sun in the sky.
Given the fact that data are taken over a certain interval of time, 
the instantaneous \PeTw\ in eq. (\ref{eq:neutrino:survival:proba}) has to be 
replaced with its time averaged, \APeTw, which requires the computation of 
the apparent solar trajectory in the detector sky.

\Item
The exact solar trajectory is a complicated function of time which changes 
with the epoch of the  year and the geographical location.
Its computation is described in Appendix A, while for the present discussion 
it is enough to recall its main features.

\Item
Looking at the Sun as at a star projected on a sphere centered on the detector 
location, its position is specified by a couple of spherical co-ordinates
\cite{Lang:1980, Nautical:Almanac:1996}. 
Assuming the Earth is a spherical and isotropical body, 
the Sun's Nadir angle is enough to describe the Sun position in order to 
compute the probability \PeTw.
The {\em Nadir angle} \hath\ is the 
angle subtended by Sun's and Earth center directions as seen from the detector 
\footnote{In a Local Reference Frame centered on the detector D 
(see Fig. \FigEarthEffectGeometry) 
the {\em Nadir} is the point on the celestial sphere which lies 
below the detector along its vertical axis, i.e., it is the point opposite 
to the detector Zenith.}, 
as depicted in Fig. \FigEarthEffectGeometry. 
During the day $\hath$\ is greater than $90\degres$\
(and $\PeTw \equiv \STvS$), while at night $\hath \le 90\degres$,
the Sun rises or sets when $\hath = 90$\degres\ and
a solar neutrino will cross the Earth center when $\hath = 0\degres$.

\Item
The Sun's apparent trajectory in the sky with respect to the detector 
is a function of the day of the year $d$. 
Each night the Sun reaches a minimum Nadir angle, $\hathmid_{,d}$. 
In the course of the year $\hathmid_{,d}$\ has a maximum value,
$\hathmmax$, and a minimum value, $\hathmmin$. 
Thus, given a point in the interior of the Earth seen (with respect to the 
detector) along a line of view of \NA\ $\hath$, it will never 
be on the solar neutrino trajectory during the night if $\hath < \hathmmin$;
it will be on a trajectory for some of the nights during the year if 
$\hathmmin \leq \hath < \hathmmax$, while if 
$\hath \ge \hathmmax$ it will be crossed by neutrinos every night.

\Item
These limits and the apparent trajectory of the Sun, are functions of the
detector location specified by its latitude, $\lambdaDtc$, and longitude.
For a detector outside the tropical band, $|\lambdaDtc| > 23\degres27'$,
\hathmmax\ is reached at summer solstice, 
\hathmmin\ is reached at winter solstice, 
and $\hathmmin > 0\degres$\ so that neutrinos received by such detectors never
cross the Earth center. 
Instead, for detectors located inside the tropical band, $\hathmmin = 0\degres$\
and neutrinos crossing the Earth center are detected, while the minimum and maximum
night \NA\ are reached at epochs between equinoxes and solstices. At 
last for an equatorial detector $\hath = 0\degres$\ is reached just at equinoxes.

\Item
It is well know that for a real time detector as \SK\ it is possible to compute
\hath\ for the Sun at the time $t$ of the detection of a given event, allowing 
the selection of events produced when the Sun is seen within a given \NA\
interval $[\hath_1, \hath_2]$. The boundaries of this interval are crossed at 
times $\hatt_1$\ and $\hatt_2$. As will be shown further, this selection may be
a feasible strategy to increase the sensitivity to the \daynight\ effect.

\Item 
It follows from the above discussion that in order to take into account the 
Earth effect in the signal collected in the time interval $[T_1, T_2]$ of the 
year by a given detector it is enough to replace \PeTw\ in eq. 
(\ref{eq:neutrino:survival:proba})
with its time averaged \APeTw\ defined as:
\bec\beq\label{eq:APeTw:Time}
\APeTw =
  \frac{1}{\TResid}
{\displaystyle    
  \int_{T_1}^{T_2} dt\, \hatdelta(\hath_1 \leq \hath(t) \leq \hath_2) \, 
       \PeTw(\hath(t)) },
\eeq\eec

\noindent
where
$[\hath_1, \hath_2]$ is the \NA\ interval of interest, 
$\hatdelta$\ is the sampling function and 
\bec\beq
  \TResid = {\displaystyle \int_{T_1}^{T_2} dt \, \hatdelta(\hath_1 
            \leq \hath(t) \leq \hath_2) }
\eeq\eec

\noindent
is the {\em residence time}, i.e., the time spent by the Sun in the 
\NA\ interval.

\Item
The values of \APeTw\ and \TResid\ are functions of the \NA\ interval,
the time interval and of the sampling scheme 
described by the sampling function.
Here it is assumed that the sampling 
i) is continuous over the time interval in which the Sun is in the relevant 
Nadir angle interval $[\hath_1, \hath_2]$,
ii) is extended over an integer set of years, 
and iii) that the error in the reconstruction of the apparent Sun
position at the time of the detection of each solar neutrino event is 
negligible. 
With these 
hypotheses $\hatdelta(\hath_1\leq \hath \leq \hath_2)$\ is $1$ when \hath\ is 
inside the interval $[\hath_1, \hath_2]$ and zero otherwise.

\Item
The choice of the \NA\ interval is quite arbitrary, but a natural one 
is to compare data collected at \DAY\ 
($\hath > 90\degres$) with data collected during the full \night\
($0\degres \le \hath \le 90\degres$),
and to separate neutrinos detected at night into \core\ neutrinos,
$0\degres \le \hath \le \hath_c$,
and \mantle\ neutrinos,
$\hath_c \le \hath \le 90\degres$,
where $\hath_c = 33.17\degres$\ is the apparent \NA\ 
of the core/mantle boundary in the Stacey model.
The three averages are labeled respectively: \APeTwC, \APeTwM\ and \APeTwN,
while the corresponding residence times are labeled \TResidC, \TResidM; 
the \night\ residence time \TResidN\ is identical to half an year 
for any detector.
The day averaged probability $\APeTw_{day}$\ coincides with \STvS.
Finally, it is possible to consider also the \deepcore\ average probability 
for which the \NA\ interval in eq. (\ref{eq:APeTw:Time})
is $\hathmmin \le \hath \le 2/3 \hath_c$.
The symbols \DAY, \night, \mantle, \core\ and \deepcore\ will be used in 
what follows to indicate these particular solar neutrino event samples.

\Item
In the absence of neutrino oscillations the number of solar neutrinos induced 
events in a given \NA\ interval is proportional to \TResid\ for that interval.
Table \TabResidenceTime\ lists a set of residence times for the \SK\ detector,
showing that in one year, for approximately 14\% of the total night time 
the Sun is seen
behind the Earth core, so that about $7\%$\ of neutrinos cross the core.
With a statistics of $10^4$ events this corresponds to a relative statistical
error of about $3.8\%$.

\Item 
The residence time is also relevant because it determines the weight of the  
contribution of each geophysical structure $s$, 
$s = $\night, \mantle, \core, \deepcore,
to $\APeTwN$, the contribution being proportional to the ratio 
$T_{Resid}^{s} / \TYear/2$, where $\TYear = 365.24$\ days.
Thus, for the \SK\ detector  the \core\ contribution in \APeTwN\ has to be 
weighted by the factor $0.07$\ and it is no longer a dominant contribution 
even when \APeTwC\ is much larger than \APeTwM.

\Item
Many types of systematic errors can affect the predictions for the \daynight\ 
asymmetry. 
Among them the detector specifications such as detection efficiency, energy 
resolution, etc. are undoubtedly very important. 
They are neglected in the present study because they are not yet exactly known
for the \SK\ detector, and moreover they may change during the various phases 
of data taking.
In this case the only experimental feature to be taken into 
account is the energy threshold, while the most relevant error sources which 
have to be discussed are the numerical errors and the Sun's apparent motion 
reconstruction.
Let us add that the error introduced by uncertainties in the detection 
cross section for the $\nue - e^-$\ elastic scattering process is less than $1\%$ 
\cite{Bahcall:Kamionkowski:Sirlin:1995} and is neglected.
The same is true also for the detector location which is known with an accuracy
better than 1 arcmin or 1 over $10^4$ \cite{Suzuki:1996}.

\Item
The choice of a correct approximation for the apparent solar motion is a 
crucial point in order to reach the required accuracy, given the Earth model 
used in the calculations.
The motion is quite complex and 
approximations are required to obtain an acceptable computation time
but at the price of a certain error. There are two kinds of approximation 
made which are relevant for the Earth effect. 
In the first the Earth motion around the Sun is 
circular (COA), in the second elliptical (EOA). Neither the first nor the 
second are exact models, but EOA  is very accurate for Earth effect prediction 
purposes.
The analysis (see Appendix  \ref{Appendix:A})
shows that most of the errors introduced by COA and EOA 
are periodic in nature, so that in the turn of one year they should  
averaged out. 
However, given that \PeTw\ is a function of the solar position and that 
the use of different sampling schemes (as \core, \mantle, \night, etc.) are 
equivalent to averaging over specific fractions of the time of the 
year only, it is not 
possible to test such approximations without accurate numerical experiments. 
These experiments show that indeed the error introduced by the simpler 
COA model can be close to (but does not exceed) $1\%$, and that a greater 
accuracy is reached by EOA. 
In this way the use of EOA for all the computations presented in this 
paper and in \cite{ArticleII}\ assures an accuracy better than $1\%$\ for 
each kind of sampling. The COA can be used when a large amount of simulations 
are required, allowing a substantial reduction of the computational time but 
at the cost of an increase in the error in \APeTw, which can reach $0.8\%$.

\Item
In the Elliptical Orbit Approximation the averaged probabilities and 
residence times defined by (\ref{eq:APeTw:Time}) have to be modified to take 
into account the change in the Sun - Earth distance with time, which causes a 
periodical change in the solar neutrino flux and in the Earth orbital velocity.
To take both effects into account it is not enough to replace the relation
between $\hath$\ and the time of the year $t$, but also the form of 
(\ref{eq:APeTw:Time}) has to be adapted.
Let \RSE(t)\ be the time dependent Sun - Earth distance expressed in units of 
the mean distance (the Astronomical Unit $R_0 = 1.4966 \times 10^8$\ km).
Then each instantaneous sample has to be weighted by the factor $1/\RSE^2(t)$\ leading to:
\bec\beq\label{eq:APeTw:Time:EOA}
\begin{array}{lll}
  \APeTw &=& 
     \frac{1}{\TResid}
    {\displaystyle \int_{T_1}^{T_2} dt \,
      \frac{\hatdelta(\hath_1 \leq \hath(t) \leq \hath_2)}{\RSE^2(t)} \,
          \PeTw(\hath(t)) },\\
&&\\
   \TResid &=& {\displaystyle \int_{T_1}^{T_2} dt \,
                \frac{\hatdelta(\hath_1 \leq \hath(t) \leq \hath_2)}
                    {\RSE^2(t)} }.\\
\end{array}
\eeq\eec

\noindent
As it is discussed in Appendix \ref{Appendix:A}, the effect of the factor 
$1/\RSE^2$\ is partially compensated by the change in the residence time due to 
the change of the Earth orbital velocity during the year.
\EndE

\section{The Instantaneous \PeTw\ Probability}\label{section:3}
\BeginE
\Item 
At the basis of each time averaged probability \APeTw\ is the instantaneous 
one, \PeTw. 
Many features of \APeTw\ are a direct consequence of those present in \PeTw.
It is natural to begin the discussion of our numerical results with those
obtained for \PeTw.

\Item
Figure \FigPeTw\ shows an example of the probability \PeTw\ for $\SdTvS = 0.01$
as a function of the {\em resonance density} \rhoR\ which is connected to $E$\ 
and the neutrino transition parameters \dms\ and \SdTvS\ through the resonance 
condition:
\bec\beq\label{eq:resonance}
      \rhoR \Ye = 6.57 \times 10^6 (\dms /\Enu) \CdTv,
\eeq\eec


\noindent
where $\rhoR$\ is expressed in gr/cm$^3$, $\Enu$ in MeV, \dms\ in eV$^2$,
\Ye\ is the electron fraction per nucleon.
We have for the Earth $\Ye = 1/2$\ with a rather high accuracy.
For a fixed \CdTv\  equation (\ref{eq:resonance}) justifies the 
use of \rhoR\ as an independent variable instead of \EDms.
The probability \PeTw\ as a function of \rhoR\ was computed solving the 
two-neutrino propagation equation \cite{LW:Barger:Langacker, 
Kuo:Pantaleone:1989b}\ using a 4$^{th}$\ order, adaptive step-size 
Runge - Kutta algorithm adapted from \cite{Num:Rec:1986}.
Each plot in Fig. \FigPeTw\ refers to a different \NA.
There are two main peaks, \calC\  (at $\rhoR \cong (10 - 11)$ gr/cm$^3$)
and \calM\ (at $\rhoR \cong (4 - 6)$ gr/cm$^3$), 
associated with resonant transitions in 
the core (\calC) and in the mantle (\calM). 
This correspondence is based on the following observations:
i) a geological structures is relevant for the neutrino transitions 
  (and therefore for $\PeTw$) when neutrinos cross it,
ii) the structure affects the neutrino propagation when its density 
    is the resonance region for the crossing neutrinos,
iii) when one of the previous conditions are not fulfilled \PeTw\ is close to 
     its vacuum value \STvS.
When \SdTvS\ increases, for instance, the resonance width also increases and it
is possible to have a rather large \PeTw\ for values of \rhoR\ which do not 
correspond to any density crossed by neutrinos along their trajectory.
In addition, the spherical symmetry of the Earth implies that solar neutrinos 
will cross the resonance region twice. This can produce interference terms 
in \PeTw\ leading to oscillatory dependence of \PeTw\ on \hath\ for a fixed 
\rhoR\ \cite{Baltz:Weneser:1994}. 
They can also be responsible for the presence of multiple peaks in \PeTw\ as 
a function of \rhoR\ for a fixed \hath,  associated with the resonances in 
a given geological structure. 
An example of this possibility is the splitting of the core peak \calC\ in 
Fig. \FigPeTw(a) in two peaks, \PeakCI\ and \PeakCO, shown in 
Fig. \FigPeTw(c).
Indeed, it is reasonable to expect the interference terms in \PeTw\ to be 
significant in certain cases because the neutrino oscillation length in matter
for solar neutrinos which undergo resonant transitions in the Earth is 
 of the order of the Earth radius.

\Item
It is instructive to see how the \PeTw\ dependence on \rhoR\ changes with the 
increasing of \SdTvS.
For $\SdTvS < 0.1$\ the qualitative properties of this dependences do not 
change significantly: \PeTw\ is basically ``rescaled'' in accordance with the 
change of \SdTvS\ and $(\PeTw - \STvS)$\ goes to zero rather quickly outside 
the resonance region. 
For larger values of \SdTvS\ the behavior is more complicate because new 
peaks appear, becoming more and more prominent as \SdTvS\ increases. 
The ``scaling'' of \PeTw\ is not precise, different parts of the plot scale
differently, as is well illustrated by the \calC\ and \calM\ peak heights. 
With the rise of  \SdTvS\ \calC\ increases 
faster than \calM\ reaching a maximum
value and begins to decrease while \calM\ is still increasing. 
As a consequence, at values of $\SdTvS \, \lsim \, 0.13$\ 
most of the \daynight\
asymmetry is generated by the MSW effect in the core.
This is not true  in the large mixing angle region.
For practical purposes the most important lesson which can be drawn from the 
\PeTw\ plots is that to produce accurate predictions for the Earth effect, 
\PeTw\ has to be computed for resonance densities \rhoR\ 
(or equivalently, for values of $E/\Delta m^2$) which are largely 
outside the density range of the Earth geological structures.
\EndE

\section{Averaging the \PeTw\ Probability}\label{section:4}
\BeginE
\Item
Figures \FigAPeTw.1 - \FigAPeTw.18 depict the results of numerical calculations
of \APeTw\ and \PTot\ for the
\SK\ detector in the case of solar $\nu_e$ transitions into an active
neutrino \footnote{See appendix \ref{Appendix:A}\ 
for further details.}, $\nu_e \rightarrow \nu_{\mu (\tau)}$.
Each figure is divided into four ``frames'' which are labeled 
(a), (b), (c) and (d).
Figures \FigAPeTw.1a - \FigAPeTw.18a display \APeTw\ for \night\ 
(short-dashed line), 
\mantle\ (dotted line), \core\ (solid line) and \deepcore\ (long-dashed line)
as a function of \rhoR\ and are the subject of the discussion in this Section.
Comparing \APeTw\ with \PeTw\ shown in Fig. \FigPeTw,
it is possible to identify the peak due to the resonance in the mantle, \PeakM,
present also in \PeTw.
Similarly, there are two very well defined peaks
(or a peak, \PeakC, with two maxima) in \APeTw, \PeakCI\ (at larger \rhoR)
and \PeakCO\ (at smaller \rhoR), 
which correspond to the resonance in the core. 

The presence of wiggles is evident for \rhoR\ greater than the core density,
especially at large mixing angles. 
These wiggles are dumped and disappear as \rhoR\ increases.
A curious feature is that in most of the cases the wiggles come in triplets 
(see Fig. \FigAPeTw.14a). 
The Log-Log plots of the \APeTw\ probability (not reported here) show that 
for \rhoR\ outside the range of the Earth density \APeTw\ decreases as follows:
\bec\beq\label{eq:eqdecrease}
    \log \left(\frac{\APeTw}{\STv^2}-1\right) = 
         A + B \log \rhoR + \epsilon f(\rhoR),
\eeq\eec

\noindent
where $A$, $B$ and $\epsilon$\ are real constants and $f(\rhoR)$\ is a
dumped oscillating function which represents the wiggles.
Since $\epsilon$\ is small ($\lsim 10^{-2}$), $f(\rhoR)$ can be neglected, 
while $A$ and $B$ can be determined by numerical means for any \APeTw. 
Equation (\ref{eq:eqdecrease}) is used to extrapolate \APeTw\ at values of 
\rhoR\ far from the Earth density.

\Item 
As Figs.\FigAPeTw.1a - \FigAPeTw.18a illustrate, the separation in \core\ and 
\mantle\ samples is very effective in enhancing \APeTw. 
The enhancement is particularly large for small mixing angles for which 
\APeTwC\ can be up to six times bigger than \APeTwM.
This suggests the possibility of a corresponding enhancement of the 
$e^{-}$-spectrum distortions
as well as of the energy integrated event rate. The latter can make feasible
the detection of the Earth effect even at small mixing angles. 
Further enhancement of the probability \APeTw\ can take place if the size of the 
core bin is decreased, as shown by the \deepcore\ averaged probability plotted 
with a long-dashed line.
However, the reduction in the bin size reduces the residence time in the 
\deepcore\ bin to less than $4\%$\ per year and it is not clear whether 
the probability enhancement can compensates the corresponding reduction
in statistics.
\EndE

\section{The Solar $\nu_e$ Survival Probability}\label{section:5}
\BeginE
\Item 
The connection between the peak structures in  
the probabilities \APeTw\ and \PTot\, considered as functions of 
$E/\Delta m^2$, is determined basically by 
eq. (\ref{eq:neutrino:survival:proba}). 
The presence of an enhancement in the averaged probability \APeTw\ 
is not enough to ensure an enhancement in the \daynight\ asymmetry. 
This is especially true at small mixing angles at which the 
Earth effect may produce reduction instead of increase of the \nue\ flux. The 
latter happens when $\Ps > 0.5$, 
so that most of neutrinos coming from the Sun are 
\nuone\ which are converted into \num, \nut\ in the Earth.

\Item 
The upper parts (``windows'') of the 
``frames'' \FrameN, \FrameC\ and \FrameM\ shown in 
Figs. \FigAPeTw.1 - \FigAPeTw.18
are combined plots of \Ps\ (dotted line), \PeTw\ 
(dashed line) and \PTot\ (solid line) for \night\ \FrameN, \core\ \FrameC\ 
and \mantle\ \FrameM\ as functions of \EDms. 
The lower ``windows'' of each ``frame'' are plots of the ratio:
\bec\beq
          \AsymP \equiv 2~\frac{\PTot - \Ps}{\PTot + \Ps}~~,
\eeq\eec

\noindent
which mimics the asymmetry used by the Kamiokande collaboration 
in their discussion of the \daynight\ effect. This quantity will 
also be called {\em asymmetry}.
In some cases \APeTw\ is too small to be visible in the full scale plot,
and so it is rescaled by a factor of $10$. This is indicated by the 
label ``\rescaled'' in the upper right corner of the corresponding figure.

\Item
The purpose of these figures is to illustrate how each probability contributes in 
\PTot\ and in the asymmetry \AsymP. 
One of the important features that has to be taken into account in the discussion 
of \PTot\ is the position of the peaks in \APeTw\ with respect to the value of 
\EDms\ at which $\Ps = 0.5$\ because it controls the sign and 
the magnitude of the \daynight\ effect.
It follows from eq. (\ref{eq:neutrino:survival:proba}) that the asymmetry is 
zero each time $\Ps = 0.5$\ irrespective of the value of \APeTw. 
If $\APeTw > \STvS$, but $\Ps > 0.5$, the asymmetry is negative which 
means the Earth effect {\em reduces} the night event rate instead of  
{\em increasing} it.
The positive Earth effects occurs when $\Ps < 0.5$, i.e., when the peak of \APeTw\ 
falls inside the lower part of the ``pit'' of the MSW probability \Ps. 
Figures \FigAPeTw.1 - \FigAPeTw.18 illustrate all these possibilities 
together with less relevant cases corresponding to $\APeTw = \STvS$\ and 
$\APeTw < \STvS$. The presence of a zero Earth effect line together with
cases in which the asymmetry \AsymP\, as a function of $E/\Delta m^2$,
can be both negative and positive, can reduce 
the magnitude of the Earth effect. 
This reduction is especially important at small mixing angles.

\Item 
The increase of \SdTvS\ within the small mixing angle solution region (3)
does not change considerably the minimum and maximum values of \Ps, but it 
affects significantly the width of the ``pit'' of \Ps. 
At the same time  the peaks in \APeTw\ change mainly their height 
and width but not
their position with respect to the \EDms\ axis. 
For $\SdTvS \le 0.002$\ most of the relevant 
part of \APeTw\ is confined in the 
$\Ps > 0.5$\ region and the Earth effect is negative. 
With the increase of \SdTvS\ the \Ps\ ``pit'' becomes wider.
Correspondingly, larger parts of the \APeTw\ peak enter 
the $\Ps < 0.5$\ region:
depending on the value of $E/\Delta m^2$, the asymmetry \AsymP\ 
(the Earth effect) can be negative or positive. 
This continues until $\SdTvS \cong 0.006$, for which most of the 
\APeTw\ peak is in the $\Ps < 0.5$\ region and the effect is mostly positive. 
This implies that even if \APeTw\ is relatively large, 
there will be a region at 
small mixing angles where the asymmetry is zero and the \core\ enhancement is
not effective. 

\Item 
The sequence of \AsymP\ plots allows to identify various peak components 
which we will denote by
\PeakA, \PeakHP, \PeakHM, \PeakA\ being the left 
one (located at smaller values of
$E/\Delta m^2$), while \PeakHP\ (\PeakHM) is located at larger
$E/\Delta m^2$ in the region where $\AsymP > 0$\ ($\AsymP < 0$). 
The \PeakA\ structure is an artifact of the presence of the adiabatic 
minimum in the probability \Ps, $min\Ps\ =\sin^2\theta_{V}$,  
which can lead to a relatively large asymmetry \AsymP\ even if
\APeTw\ is small. 
The smallness of \APeTw\ in the \calA\ region means that 
that this region does not contribute significantly 
to the total event rate asymmetry.

\Item 
The bulk of the \daynight\ asymmetry comes 
from \PeakHM\ and \PeakHP\ structures, 
which are directly connected to the \PeakC\ and \PeakM\ peaks in \APeTw. 
For $\SdTvS \le 0.002$\ (Figs. \FigAPeTw.1 - \FigAPeTw.3) most of the peaks in 
\APeTw\ lie in the $\Ps > 0.5$\ region. 
Correspondingly, the asymmetry \AsymP\ is negative (only the peak 
\PeakHM\ is present). 
It increases in absolute value as \SdTvS\ increases and it is sensitive to the 
\core\ enhancement. 
As \SdTvS\ grows further, the bulk of the \APeTw\ resonance part enters the region 
where $\Ps = 0.5$, producing a suppression of the asymmetry and of the \core\ 
enhancement, as is illustrated for $\SdTvS = 0.004$\ in Fig. \FigAPeTw.4. 
It is interesting to note that this crossing occurs at different \SdTvS\ for 
the \night, \core\ and \mantle\ samples.
As is shown in Fig. \FigAPeTw.4\FrameC,
the peak \PeakC\ for the \core\ is nearly cut in
two parts by the $\Ps = 0.5$\ line. This leads to
the presence of both positive and 
negative Earth effect in the asymmetry \AsymP,
while in the cases of \night\ and \mantle\ (Figs. \FigAPeTw.4\FrameN\ and 
\FigAPeTw.4\FrameM) the asymmetry is negative.
At $\SdTvS = 0.006$\ (Fig. \FigAPeTw.5) the peak \PeakC\  
is in the $\Ps < 0.5$\ region while the bulk of the effect of the \mantle\ 
is still in the $\Ps \ge 0.5$\ region. 
The \core\ asymmetry is positive and it has a maximum value of 
$16.4\%$.  For $\SdTvS \ge 0.008$\ (Figs. \FigAPeTw.6 - \FigAPeTw.18)
the effect is positive.
All these details are influenced by the position of the 
$\Ps = 0.5$\ line with respect to the region of the \APeTw\ maxima. 
Given the fact that the latter is primarily a function of the Earth model, 
it cannot be excluded that certain aspects of the behavior described above  
may change somewhat with the change of 
the Earth model used in the calculations.


\Item 
Finally, the sequences of figures \FigAPeTw\ show 
that the resonance in the core 
is the dominant source of the asymmetry in the region of the 
nonadiabatic solution, $\SdTvS \leq 0.013$\, while for 
$\SdTvS \geq 0.3$\ the bulk of the asymmetry is generated 
by the resonance in the mantle (the \PeakM\ 
peak) and the \core\ enhancement is only of the order of $30 \div 40\%$.
\EndE

\section{Conclusions}\label{section:6}
\BeginE
\Item
In the present article and in  ref. \cite{ArticleII}\ we 
have performed a detailed study of the \daynight\ effect 
for the \SK\ detector assuming the MSW solution of the 
solar neutrino problem corresponding to solar \nue\ 
transitions into an active neutrino, $\nue 
\rightarrow \numt$.
Among our aims were:
i) calculation of the Earth effect with a sufficiently high
precision which can match the precision of the data on the effect to be 
provided by the \SK\ detector, 
ii) calculation of the magnitude of the effect 
for a sufficiently large and representative set of values of the 
parameters  from the ``conservative'' regions (\ref{eq:kpnu96:aII}) and 
(\ref{eq:kpnu96:bII}) of the MSW solution, 
iii) calculations of the one-year
averaged D-N asymmetry for solar neutrinos crossing the Earth mantle, 
the core, the inner 2/3 
of the core, and mantle + core (full Earth) for the chosen large set of 
values of \dms\ and \SdTvS\ with the purpose of illustrating the 
magnitude of the enhancement of the effect which can be achieved by an 
appropriate selection of the data sample, 
iv) calculation of the one-year average deformations of the $e^-$\ 
spectrum by the Earth effect in the mantle, the core, and in the 
mantle + core 
for the selected representative set of values of \dms\ and \SdTvS. 
In the present article we have used the elliptical orbit approximation for the
Earth motion around the Sun to calculate the one year average solar \nue\
survival probability and the \daynight\ asymmetry 
in the probability for neutrinos crossing the 
Earth mantle, the core, the inner 
2/3 of the core, and the mantle + core.
The indicated sampling which can be done with the \SK\ detector, 
was considered in order to investigate quantitatively 
the possibility of enhancement of the \daynight\ effect. 
We have found, in particular, that such an enhancement can be especially large
at small vacuum mixing angles, $\SdTvS \lsim 0.013$: 
the \daynight\ asymmetry in the one-year averaged 
\nue\ survival probability for neutrinos 
crossing the Earth core only can be a factor of up to six bigger 
than the asymmetry in the analogous probability for 
neutrinos crossing the core + mantle
(\night\ data sample).
This result persists in the corresponding event rate 
samples \cite{ArticleII},
thus suggesting that it may be possible to test the $\SdTvS \le 0.01$\ region 
of the MSW solution of the solar neutrino problem 
by performing selective \daynight\ 
asymmetry measurements with the \SK\ detector.

We have studied also the accuracy of the circular orbit approximation (COA) 
utilized in many of the previous studies of the \daynight\ effect, and compared
it with the accuracy of the elliptical orbit approximation (EOA) we have used 
in the present analysis.
The main conclusion of this study is that the 
largest difference in the results for the probability 
\APeTw\ obtained using the COA and EOA is $0.8\%$.
Given the fact that the EOA was chosen for the computations in this work, 
and that the EOA program actually includes most of the secondary celestial 
mechanics effects, the accuracy of the predictions 
for \APeTw\ in what concerns the motion of the Sun 
is better (and probably much better) than $1\%$.

\section*{Acknowledgments}\label{section:7}
M.M. wishes to thank the International School for Advanced Studies, Trieste, 
Italy, where part of the work for this study has been done, for kind 
hospitality and financial support. 
The authors are indebted to the ICARUS group of the University of Pavia 
and INFN, Sezione di Pavia, and especially to Prof. E. Calligarich, 
for allowing the use of their computing facilities for the present study.
M.M. wishes to thank also Dr. A. Rappoldi for his suggestions 
concerning the computational aspects of the study, and to Prof. A. Piazzoli 
for his constant interest in the work and support.
The work of S.T.P. was supported in part by the EEC grant ERBFMRXCT960090
and by Grant PH-510 from the Bulgarian Science Foundation.

\vspace{0.5cm}

\appendix
\section*{Appendix: The Solar Motion Approximations}\label{Appendix:A}
The calculation of the probability \APeTw\ with a sufficient accuracy
requires a relatively good approximation for the 
Sun's apparent motion which to leading order can be considered as a 
superposition of two strictly periodical motions, the orbital revolution and 
the Earth rotation, plus a set of small perturbations.
The exact solution of such perturbation problem is complex and 
computationally demanding and therefore one has to select only those effects 
which are relevant for the chosen accuracy threshold $\epsilon$\
(ref. \cite{Lang:1980, Zagar:1984, AOPC, Nautical:Almanac:1996}).
The various errors introduced by inadequate approximations are systematic in 
nature. They combine algebraically leading to subtle cancelation or enhancement
effects which are difficult to propagate to the probability 
\APeTw\ of interest.
However, a practical approach to cut off all corrections which are not 
relevant is to set a threshold in the Sun's motion 
tracking accuracy of the order of some arcmin.
This threshold may be justified by the fact that the Sun is not a point like 
neutrino source and the neutrino production region has an apparent angular 
diameter of about 3 arcmin (approximately one tenth of the solar diameter).
The correctness of the indicated choice is confirmed by a 
more detailed analysis.

 Using the accuracy threshold it is possible to discriminate between the 
various assumptions which can be used to simplify the apparent solar 
motion. As a result of such an analysis it becomes evident that two orbital 
approximations are sufficiently accurate for our purposes:
the circular orbit approximation (COA) and 
the Elliptical Orbit Approximation (EOA).
In the COA the Sun's orbit is circular, while in the EOA the 
true elliptical Sun's orbit
described by Kepler's laws is considered.
There are two important effects which are 
neglected by COA and which instead are included in EOA: 
i) the time dependence of the Sun - Earth distance, $\RSE(t)$, during the year,
   which leads to a change in the solar neutrino flux at the Earth, 
$\Phi$, according to the $1/\RSE^2(t)$\ law,
ii) the time dependence of the Earth orbital velocity, $v(t)_\oplus$, during 
    the year due to the angular momentum conservation.
The velocity $v(t)_\oplus$
    also changes according to the $1/\RSE^2(t)$\ law. This implies 
    a change in the residence time of the Earth, $\TResid_{,\oplus}$,
    in a given orbital location, which is  
proportional to $\RSE^2(t)$.
Each of these two effects can introduce a systematic error of up to $3\%$.
When taken together into account they compensate partially each other
because the total neutrino rate 
for a given position of the Earth with respect to the Sun
is proportional to the product 
$\TResid_{,\oplus} \Phi$ and therefore the total effect has to be 
rather small for a one year average \night\ sample.
However, when other samples as \core\ or \deepcore\ are considered  
cancelation at the $1\%$\ level cannot be guaranteed by pure heuristic 
arguments since even with an one year data taking the \core\ and \deepcore\ 
sampling are equivalent to averaging over a fraction of the year (winter).
In addition, for detectors located at positive 
latitudes over the tropical band, the \core\ sample 
is detected at winter when $\Phi_\odot$\ and $v(t)_\oplus$\ are 
maximal.

For the aforementioned reasons a set of numerical 
simulations were performed using both
COA and EOA approximations and comparing the results for both 
\APeTw\ and \TResid\ obtained with them.
As already stated in the text, the main conclusion of this study 
is that the largest difference in the results for 
\APeTw\ obtained utilizing the COA and EOA is  $0.8\%$, and that the 
3 arcmin error threshold provides a sufficient accuracy in the 
description of the solar motion.
Let us note also that the program which make use of the elliptical orbit
approximation takes into account many other celestial mechanics 
secondary effects as well, so that 
its accuracy is better than few tenth of an
arcmin. Since all the results presented in this paper are obtained with
this program, it is evident that the accuracy of \APeTw\ predictions here 
presented is far better than $1\%$.

A number of tests where also performed to assure that the accuracy of the 
numerical calculations is sufficiently high.
For instance, calculated solar positions were compared with positions given
in the 1996 edition of the {\em Nautical Almanac} 
\cite{Nautical:Almanac:1996}.
These tests showed that the numerical errors are not greater than 
$10^{-3}\%$\ and therefore are negligible. 

Finally, let us note that  eq. (\ref{eq:APeTw:Time}) can also be written  
using the ratio 
\bec$
      \frac{1}{\TResid(\hath \leq 90\degres)} \frac{d \TResid}{d\,\hath},
$\eec

\noindent
i.e., the probability to receive a neutrino from the Sun (of any flavour or 
energy) when it is inside the \NA\ interval $\hath, \hath+d\hath$.
In this case the expression for the average transition probability for 
the \NA\ interval [$\hath_1, \hath_2$] becomes:
\bec$\label{eq:APeTw:Elong}
~~~~~~~~~~~~~~~~~~~~~~~~~~~\APeTw = \int_{\hath_1}^{\hath_2} d\hath
    \PeTw(\hath) 
    \frac{1}{\TResid(90\degres)} \frac{d \TResid}{d\,\hath}~~. 
\mbox{~~~~~~~~~~~~~~~~~~~~~~~~{(A.1)}}
$\eec

\noindent
Although it looks more attractive theoretically, 
this formulation is unpractical due to the 
presence of 365 singularity points in the $d\TResid/d\hath$\ function,
as is illustrated in Fig. \FigForest. 
These singularities are due to the fact that when the Sun's \NA\ reaches its 
minimum value, $d\hath(t)/dt$\ vanishes while 
$d\TResid/d\hath \approx 1/(d\hath(t)/dt)$\ diverges. 
Obviously, these singularities are integrable but they make rather 
complicated the use of eq. (A.1)
for sufficiently accurate calculations.

\newpage


%
%
\bec{\Large\bf{Tables}}\eec\vspace{1cm}

\bec
\begin{tabular}{|l|cccc|}
\multicolumn{5}{c}{\large {Table \TabActiveNeutrinoDots.} 
            List of the $\Delta m^2$ and $sin^22\theta$ sample values}\\
\hline
       & \multicolumn{4}{c|}{\dms (eV$^2$)}\\
\SdTvS & \multicolumn{1}{c}{I}    &
           \multicolumn{1}{c}{II}    & 
             \multicolumn{1}{c}{III}    &
                   IV \\
\hline
\hline
0.0008 & 9E-6 & 7E-6 & 5E-6  &\\
0.0010 & 9E-5 & 7E-6 & 5E-6  &\\
0.0020 & 1E-5 & 7E-6 & 5E-6  &\\
0.0040 & 1E-5 & 7E-6 & 5E-6  &\\
0.0060 & 1E-5 & 7E-6 & 5E-6  &\\
0.0080 & 1E-5 & 7E-6 & 5E-6  &\\
0.0100 & 7E-6 & 5E-6 &       &\\
0.0130 & 5E-6   &        &        &\\
0.3000 & 1.5E-5 & 2.0E-5 & 3.0E-5 & 4.0E-5 \\
0.4800 & 3E-5 & 5E-5     &        &\\
0.5000 & 2E-5 &          &        &\\
0.560  & 1E-5 &          &        &\\
0.600  & 8E-5 &          &        &\\
0.700  & 3E-5 & 5e-5     &        &\\
0.770  & 2E-5 &          &        &\\
0.800  & 1.3E-4 &        &        &\\
0.900  &    &            &        &\\
\hline
\end{tabular}
\eec

\vspace{1cm}

%
%
\begin{center}
\begin{tabular}{|l|cc|c|c|c|}
\multicolumn{6}{c}{{\large{{Table {\TabResidenceTime}.}}Residence 
                        times for the \SK\ detector.}}\\ 
\hline
&\multicolumn{2}{|c|}{Nadir Angle \hath} &&&\\
\multicolumn{1}{|c|}{Sample $s$} &
  \multicolumn{1}{c}{from} &
   \multicolumn{1}{c|}{to} &
   \multicolumn{1}{c|}{$\frac{T_{res}^s}{\TYear}\times 100$} &
      \multicolumn{1}{c|}{$\frac{T_{res}^s}{\TResidN}\times 100$} &
    \multicolumn{1}{c|}{$\frac{\sigma(\Re)}{\Re} \times 100$}\\
\hline\hline
\deepcore\ & 0\degres     & 25.98\degres & 3.94  &  7.86 & 5.0     \\
\core\     & 0\degres     & 33.17\degres & 7.11  & 14.18 & 3.8        \\
\mantle\   & 33.17\degres & 90\degres    & 43.03 & 85.82 & 1.5                \\
\night\    & 0\degres     & 90\degres    & 50.14 & \multicolumn{1}{c|}{-} & 1.4 \\
\hline\hline
\multicolumn{6}{|l|}{{\bf{Note:}}}\\
\multicolumn{6}{|l|}{\TYear = 365.243 days.}\\
\multicolumn{6}{|l|}{The integration is extended over one year from 
                   day 0 to day 365}\\
\multicolumn{6}{|l|}{$\sigma(\Re)$\ is the 1 year mean 
        event rate statistical uncertainty 
                   for 
                  $\ReZr = 10^4$\ $\nu$/year}\\
\hline
\end{tabular}
\end{center}

\newpage



\newpage

\bec{\Large\bf{Figure Captions}}\eec

\noindent
{\bf Figure \FigEarthEffectGeometry.}
The geometry of the Earth effect:
{\large D} and $\odot$\ denote the detector and the Sun, 
\hath\ is the Sun's Nadir angle,
\REarth\ is the Earth radius (6371 km) and 
$R_{core}$\ is the core radius (3486 km). 
The small arrows point in the directions of the detector Zenith and Nadir.

\vspace{1cm}

\noindent
{\bf Figure \FigPeTw.}
The dependence of the probability \PeTw\ on \rhoR\ for 
$\SdTvS = 0.01$. The five plots are obtained for 
$0.1 \mbox{gr/cm}^3 \leq \rhoR \leq 30.0~\mbox{gr/cm}^3 $
and five different solar neutrino trajectories in the Earth:
a) center crossing ($\hath = 0\degres$),
b) winter solstice for the \SK\ detector ($\hath = 13\deg$),
c) half core for the \SK\ detector ($\hath = 23\degres$),
d) core/mantle boundary ($\hath = 33\degres$),
e) half mantle ($\hath = 51\degres$).

\vspace{1cm}

\noindent
{\bf Figure \FigAPeTw.} 
The probabilities \APeTw\ and \PTot\ as functions of \EDms.
Each figure represents \Ps, \APeTw, \PTot\ and \AsymP\ for the value of \SdTvS\
indicated in the figure.
The frames \FrameP\ show the probability \APeTw\ as a function 
of the resonance density \rhoR\ for the 
\night\ ({\em short - dashed Line}),
\mantle\ ({\em dotted line}),
\core\ ({\em full line})
and \deepcore\ ({\em long - dashed line}) samples.
The frames \FrameN, \FrameC\ and \FrameM\ represent the probabilities 
\Ps\ ({\em Dotted Line}), 
\APeTw\ ({\em Dashed Line}), \PTot\ ({\em Full Line})
corresponding to the \night, \core\ and \mantle\ samples as functions of 
$\Enu/\dms$.
In some cases \APeTw\ is multiplied by a factor of $10$, which is indicated 
in the corresponding figures.

\vspace{1cm}

\noindent
{\bf Figure \FigForest.}
Distribution of the differential residence time $d \TResid/d\hath\degres$\
for the \SK\ detector. The \NA\ is expressed in degrees.
The upper figure depicts the distribution for $\hath \le 89\deg$,
while the lower figure shows an enlargement of the distribution for 
$40\deg \leq \hath \leq 44\deg$.
The lower figure shows also that the small wiggles in the upper figure
are due to a set of singularities not well sampled by the computing algorithm. 
The peaks in the lower figure have a finite height only because of limitations
in the numerical computation accuracy.


\begin{thebibliography}{200}


\bibitem{MSW:Original}{
S.P. Mikheyev and A.Yu. Smirnow,
Sov. J. Nucl. Phys. {\bf{6}}, 913 (1985);\\
L. Wolfenstein, Phy. Rev. D {\bf{20}}, 2369 (1978).\\
}

\bibitem{Krastev:Petcov:1996}{
P.I. Krastev and S.T. Petcov, 
Reported by S.T. Petcov at the ``Neutrino'96'' International Conference on
Neutrino Physics and Astrophysics, June 13-19, 1996, Helsinki, Finland.}

\bibitem{Bahcall:Pinsonneault:1995}{
J.N. Bahcall and M.H. Pinsonneault,
Rev. Mod. Phys. {\bf{67}}, 781 (1995)}

\bibitem{Bahcall:Pinsonneault:etal:1996}{
J.N. Bahcall et al., 
Preprint in:  http://babbage.sissa.it, astro-ph/9610250 (1996)\\
}

\bibitem{MSW2}{
P.I. Krastev and A.Yu. Smirnov, Phys. Lett. B {\bf{338}}, 282 (1994);
V. Berezinsky. G. Fiorentini and M. Lissia, {\it ibid.} B{\bf 341}, 38 (1994);
N. Hata and P. Langacker, Phys. Rev. D{\bf 52}, 420 (1995).
}

\bibitem{Cribier:etal:1986}{
M. Cribier et al.,
Phys. Lett. B {\bf{182}}, 89 (1986);
%
J. Bouchez et al.,
Z. Phys. C {\bf{32}}, 499 (1986).
}

\bibitem{MS:DNEFF:1987}{
S.P. Mikheyev and A.Yu. Smirnov, in 
{\em New and Exotic Phenomena}, 
Proceedings of the Moriond Workshop, Les Arc, Sovoie, France, 1987, 
edited by O. Fackler and J. Tran Thanh Van 
(Editions Fronti{\`e}res, Gif-sur-Yvette, France, 1987), p. 405.
}

\bibitem{Carlson:1986}{
E.D. Carlson,
Phys. Rev. D {\bf{34}}, 1454 (1986).
}

\bibitem{Hiroi:etal:1987}{
S. Hiroi et al.,
Prog. Theor. Phys. {\bf{78}}, 1433 (1987).
}

\bibitem{Baltz:Weneser:1987}{
A.J. Baltz and J. Weneser,
Phys. Rev. D {\bf{35}}, 528 (1987);
{\it ibid.} D {\bf{37}}, 3364 (1988).
}

\bibitem{Dar:Mann:1987}{
A. Dar and A. Mann,
Nature {\bf{325}}, 790 (1987).
}

\bibitem{Baltz:Weneser:1994}{
A.J. Baltz and J. Weneser,
Phys. Rev. D {\bf{50}}, 5971 (1994).
}

\bibitem{Hata:Langacker:1993Earth}{
N. Hata and P. Langacker,
Phys. Rev. D {\bf{50}}, 632 (1994)
}

\bibitem{Krastev:1996}{
P.I. Krastev, 
Preprint in: http://www.babbage.sissa.it, hep-ph/9610339 (1996)
}

\bibitem{Gelb:Kwong:Rosen:1996}{
J.M. Gelb, W. Kwong and S.P. Rosen,
Preprint in:  http://www.babbage.sissa.it, hep-ph/9612332 (1996).
}

\bibitem{Kamiokande:Results}{
K.S. Hirata, et al.,
Phys. Rev. D {\bf{44}}, 2241 (1991);
K.S. Hirata, et al.,
Phys. Rev. Lett. {\bf{66}}, 9 (1991);
Y. Fukuda et al.,
Phys. Rev. Lett. {\bf{77}}, (1996) 1683.
}

\bibitem{Ermilova:Tsarev:Chechin:1986}{
V.K. Ermilova, V.A. Tsarev and V.A. Chechin,
JETP Letters {\bf{43}}, 453 (1986).\\
}

\bibitem{LoSecco:1993}{
J.M. LoSecco,
Phys. Rev. D {\bf{47}}, 2032 (1993).\\
}

\bibitem{Krastev:Petcov:1988E}{
P.I. Krastev and S.T. Petcov,
Phys. Lett. B {\bf{207}}, 64 (1988).
}

\bibitem{Dar:Mann:Melina:Zajfman:1987}{
A. Dar et al.,
Phys. Rev. D {\bf{35}}, 3607 (1987).\\
}

\bibitem{Nicolaidis:1988}{
A. Nicolaidis,
Phys. Lett. B {\bf{200}}, 553 (1988).
}

\bibitem{Bertotti:Maris:1996}{B. Bertotti and M. Maris, 
    {\it Variabilit\a Temporale  del Flusso di Neutrini Solari in Esperimenti 
           Sotterranei}
    ({\small{\it Neutrino Flux Time Variability in  Underground
       Experiments}}), 1996,
    reported in the 
    Ph.D. thesis of  M. Maris, Department of  
      Theoretical and Nuclear Physics (D.F.N.T.),
        Pavia University, Italy, in Italian.
}

\bibitem{Stacey:1977}{
F.D. Stacey,
 {\it Physics of the Earth, 2$^{nd}$ edition},
John Wiley and Sons,
London, New York,
1977.
}

\bibitem{Jeanloz:1990}{
R. Jeanloz,
Annu. Rev. Earth Planet. Sci. {\bf{18}}, 356 (1990).
}

\bibitem{Petcov:1988}{S.T. Petcov, Phys. Lett. B {\bf{200}}, 373 (1988).}

\bibitem{Krastev:Petcov:1988}{
P.I. Krastev and S.T. Petcov, Phys. Lett. B {\bf{205}}, 84 (1988).
}

\bibitem{Lang:1980}{
K.R. Lang,
{\it Astrophysical Formulae, 2$^{nd}$ edition},
Springer - Verlag,
Heilderberg, 1980.
}

\bibitem{Bahcall:Kamionkowski:Sirlin:1995}{
J.N. Bahcall, M. Kamionkowski and A. Sirlin,
Phys. Rev. D {\bf{51}}, 6146 (1995).
}

\bibitem{Suzuki:1996}{Y. Suzuki,  Private communication (1996).}

\bibitem{ArticleII}{
M. Maris and S.T. Petcov,
SISSA Report ref. SISSA 17/97/EP, January 1997.
}

\bibitem{LW:Barger:Langacker}{V. Barger et al., 
Phys. Rev. D {\bf{17}}, 2718 (1980);
P. Langacker et al., {\em{ibid.}} D {\bf{27}}, 1228 (1983).}

\bibitem{Kuo:Pantaleone:1989b}{
T.K. Kuo and J. Pantaleone,
Rev. of Mod. Phy. {\bf{61}}, 937 (1989).
}

\bibitem{Num:Rec:1986}{
W.H. Press, B.P. Flannery, S.A. Teukolvsky, W.T. Vetterling,
{\it Numerical Recipes},
Cambridge University Press,
Cambridge,  1986.
}

\bibitem{Zagar:1984}{
F. Zagar,
{\it Astronomia Sferica e Teorica}
({\small{\it{Spherical and Theoretical Astronomy}}}),
Zanichelli,
Bologna, 1984.
}

\bibitem{AOPC}{
O. Montembruck, T. PFleger,
{\it Astronomy on the Personal Computer},
Springer - Verlag,
Berlin, Heidelberg, 1989.
}

\bibitem{Nautical:Almanac:1996}{
{\it The Nautical Almanac}, 
Edited by the Nautical Almanac Office, Washington USA (1996). 
}


\end{thebibliography}
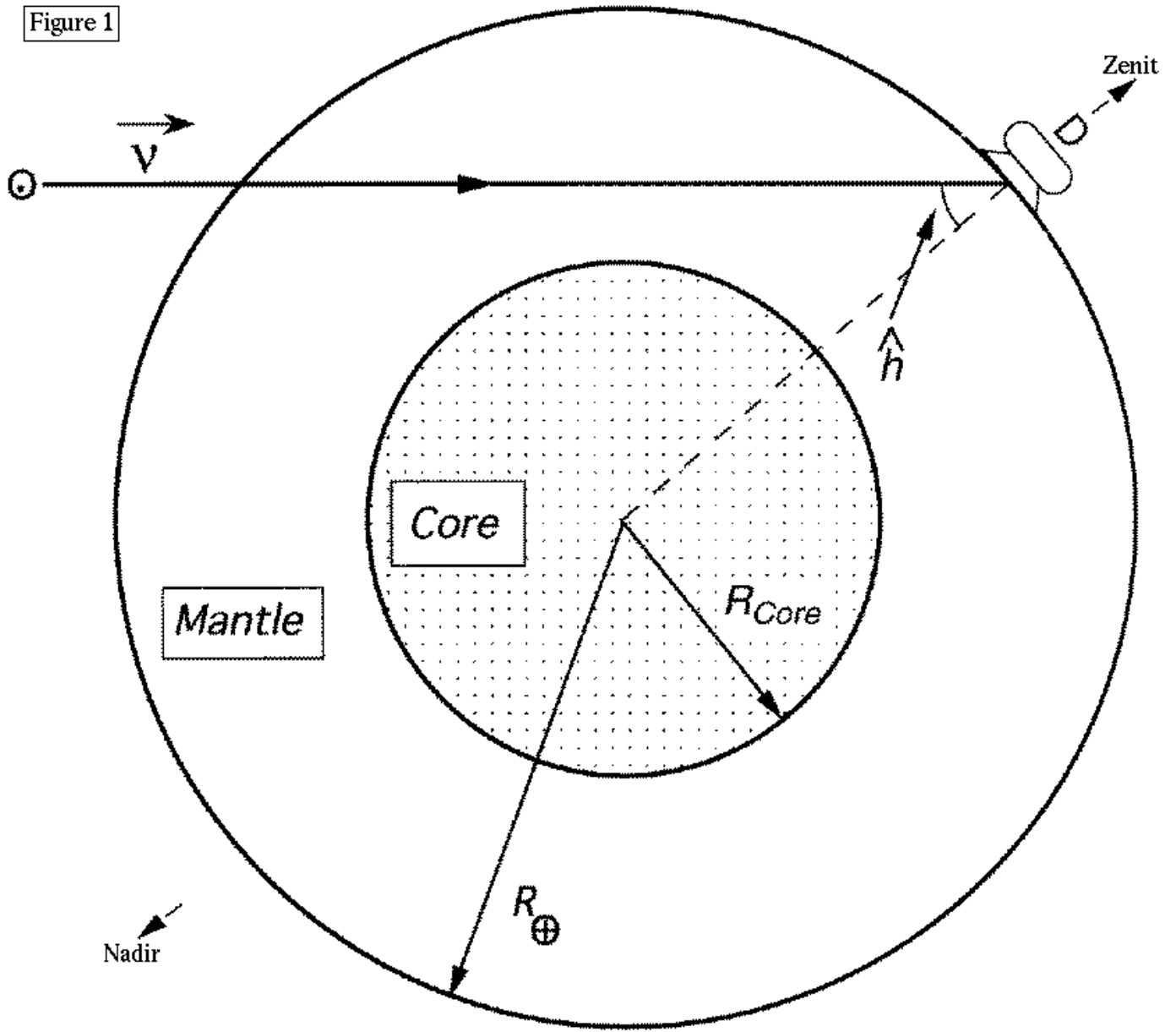
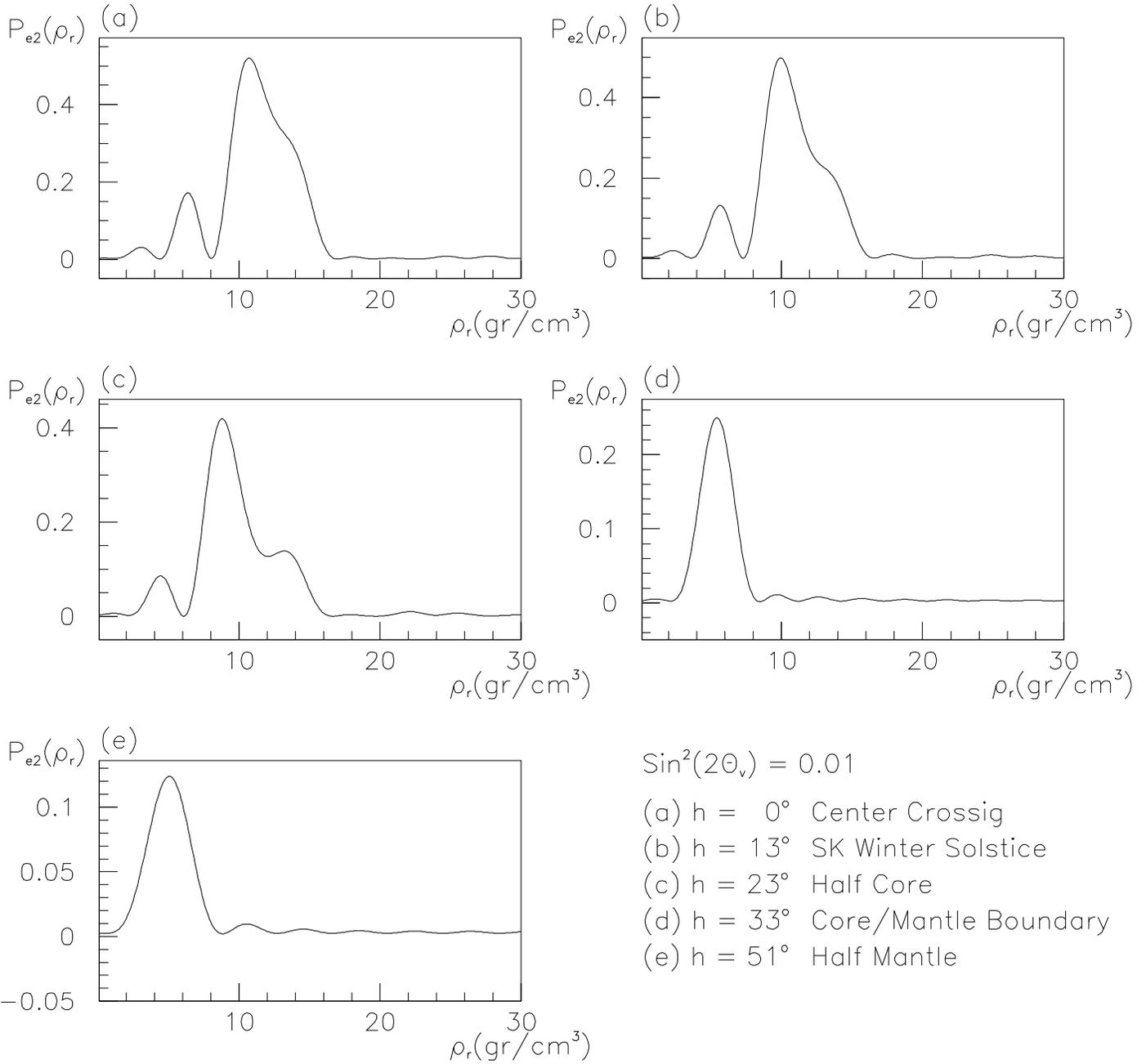
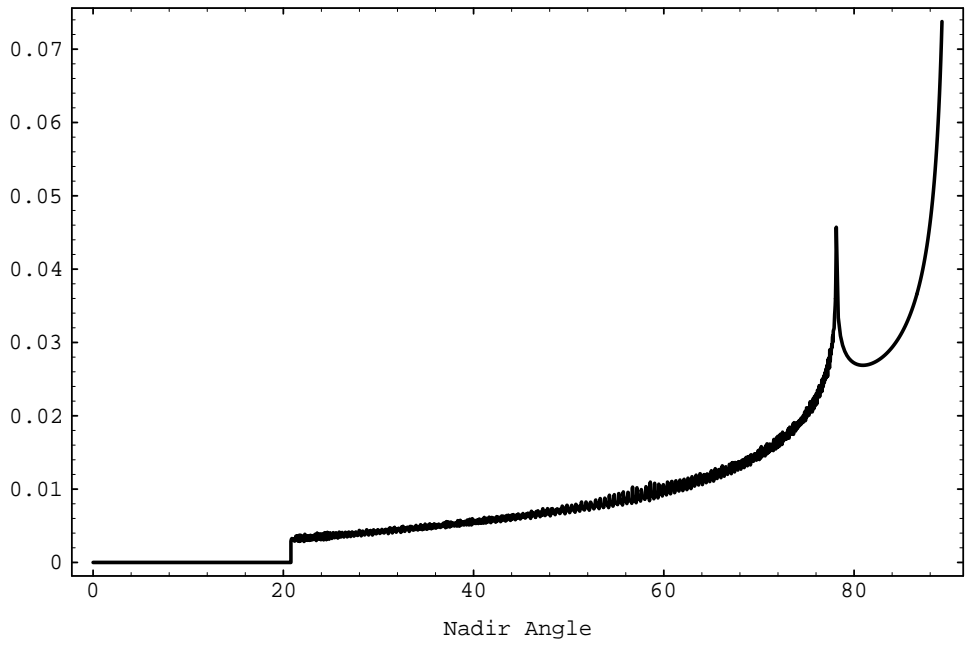
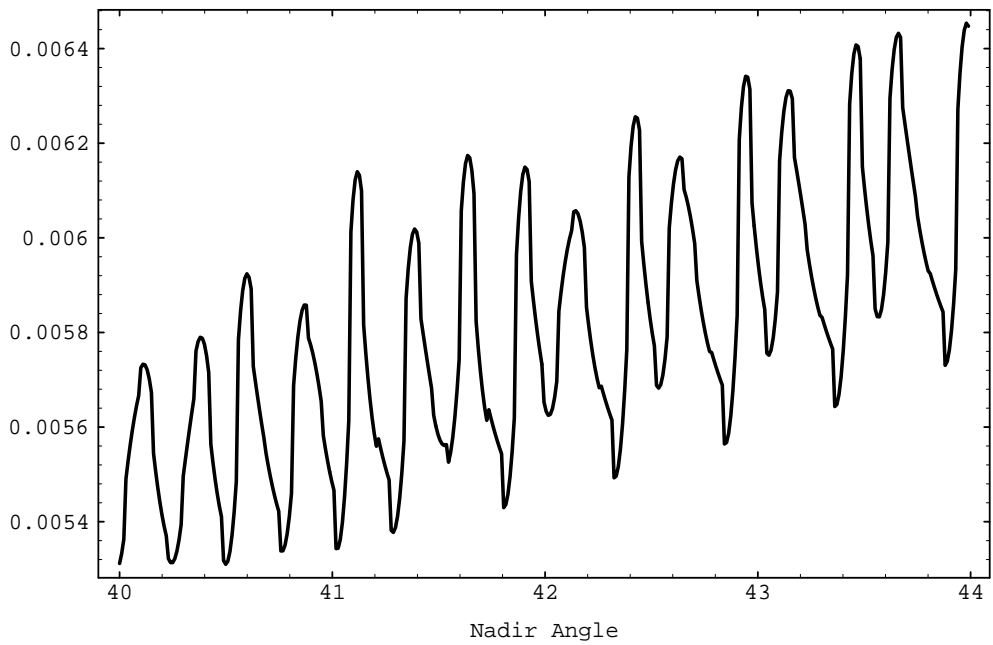
\end{document}